\documentstyle[12pt]{article}
\newcommand{\p}{{\vec p}}
\newcommand{\q}{{\vec q}}

\newcommand{\la}{\label}

\newcommand{\be}{\begin{equation}}
\newcommand{\ee}{\end{equation}}
\newcommand{\ba}{\begin{eqnarray}}
\newcommand{\ea}{\end{eqnarray}}
\newcommand{\bastar}{\begin{eqnarray*}}
\newcommand{\eastar}{\end{eqnarray*}}
\begin{document}

\begin{titlepage}

\begin{center}
\vskip 4.0cm
{\bf \large ANALYTICAL RESULTS \\ \vskip 0.3cm
            FOR COLD ASYMMETRICAL FERMION SUPERFLUIDS \\ \vskip 0.3cm
            AT THE MEAN-FIELD LEVEL \\}
\end{center}

\vskip 2.0cm

\begin{center}
{\bf Wang-Chang Su$^{*}$ }  \\
\vskip 0.3cm
{\it Department of Physics \\
     National Chung-Cheng University, Chia-Yi, Taiwan } \\
\end{center}

\vskip 0.5cm

{\rm
We present the analytical results at the mean-field level for the asymmetrical fermion system with attractive contact interaction at the zero temperature.
The results can be expressed in terms of linear combinations of the elliptic integrals of the first and second kinds.
In the limit of small gap parameter, we discuss how the asymmetry in fermion species affects the phases of the ground state.
In the limit of large gap parameter, we show that two candidate phases are competing for the system's ground state.
The Sarma phase containing a pure Fermi fluid and a mixed condensate is favored at large degree of asymmetry.
The separated phase consisting of a pure Fermi fluid and a boson condensate supports the system at smaller degree of asymmetry.
The two phases are degenerate in the limit of infinite pairing gap.
}

\vskip 1.0cm

\begin{flushleft}
PACS number(s): 03.75.Hh, 05.30.Fk, 74.20.-z
\end{flushleft}

\noindent\vfill

\begin{flushleft}
\rule{5.1 in}{.007 in} \\
$^{*}$ \hskip 0.2cm {\small  E-mail: \bf suw@phy.ccu.edu.tw } \\
\end{flushleft}

\end{titlepage}

\section{\la{Introduction} Introduction}

The recent experimental advances on the cold Fermi atoms \cite{kinast}-\cite{greiner} have revived a great amount of theoretical interest in superfluid fermion systems.
From the standard Bardeen Cooper Schrieffer (BCS) theory of superconductivity \cite{bcs}, it is known that a Fermi system with an arbitrarily weak attractive interaction induces a pairing mechanism, which takes place between the spin-up and spin-down particles with equal and opposite momenta on the common Fermi surface.
The system, being in the fully gapped BCS state, exhibits $s$-wave superfluid properties and opens a homogeneous gap in the excitation spectrum.
However, when the system is placed in the presence of a weak magnetic field, the Cooper pairing mechanism becomes unstable due to the mismatch of Fermi surfaces of the opposite spin particles.
The pairing is now likely to occur at a nonzero total momentum in the so-called Larkin, Ovchinnikov, Ferrel, and Fulde (LOFF) phase, where the gap parameter acquires a spatial variation and breaks translational symmetry \cite{loff}.

Fermion systems with mismatched Fermi surfaces and still having spherical symmetric interaction can be found in the models of asymmetrical two-component fermions, for instance, when we attempt to pair fermions with different species or attempt to pair the same fermion at finite polarization.
A state of this type was first studied by Sarma long time ago \cite{sarma}.
The state is said in the Sarma phase.
Because of the two distinct Fermi surfaces, the Sarma phase contains the phase separation in momentum space, including the paired region about the Fermi surfaces and the unpaired region in between the Fermi surfaces.
As a result, it is frequently termed the breached pair superfluid state (BP) \cite{wilczek.0}.
Besides, the same state is sometimes called the gapless superconducting phase, since a single quasiparticle dispersion gets across the free Fermi surface \cite{wilczek.1}.
In general, the gapless state of matter is characterized by the spatially homogeneous composition of the paired fermions, forming a superfluid, and the unpaired fermions, retaining the property of a Fermi liquid.

In the weak negative coupling regime, it is recently pointed out that the spatially homogeneous Sarma phase (the BP state) is unlikely to be the favored ground state for the asymmetrical fermion system \cite{caldas}.
Here, the weak-coupling regime is relevant to the limit of small gap parameter of the system.
In the context of grand canonical ensemble, on one hand, it is shown \cite{caldas} that either the normal or the BCS states is the energetically preferable ground state, depending on the degree of asymmetry between the Fermi surfaces of the two species of fermions.
On the other hand, the spatially inhomogeneous mixed phase consisting of BCS superfluid and normal states is suggested to be the true ground state in the context of canonical ensemble. Despite this, it does not rule out the possibility of constructing models in which the ground states are indeed characterized by a spatially homogeneous state coexisting superfluid and normal fluid.
One model exhibiting momentum dependent interaction with a large mass ratio is proposed to realize the stable Sarma phase (BP) \cite{wilczek.2}.
Another model, which retains spatially homogeneous property but breaks rotational symmetry, is discussed in \cite{wilczek.3} using the anisotropic pairing interaction in the weak $p$-wave coupling.

In the strong-coupling regime, {\it i.e.}, the moderate values of gap parameter, the search for the ground states of the asymmetrical fermion system is also reported \cite{carlson,pao}.
It is shown in the former reference \cite{carlson} that a homogeneous superfluid phase with a small polarization (or a small asymmetry in the fermion species) and nontrivial gapless excitations may be favored as the ground state over the inhomogeneous mixed phase.
Furthermore, at infinite coupling constant, the mixed and homogeneous phases are actually found degenerate.
That is, the gapless superfluid can exist at very strong coupling and the state may be accessible in cold atom experiments.
The findings are consistent with the results shown in the article \cite{viverit}.

The purpose of the present paper is to provide an analytical study on the asymmetrical fermion system, in order to gain clear insight on the spatially homogeneous Sarma phase in both small and large pairing gap regimes.
The system we shall consider is described by the Hamiltonian of two species of fermions interacting through an attractive contact interaction at the zero temperature.
Based on the mean-field analysis, a number of physical quantities of the system, such as the renormalized coupling constant, the thermodynamic potential, and the Helmholtz free energy, are formulated into of several integral functions.
After performing the integration, we are able to express the quantities in terms of linear combinations of the elliptic integrals of the first and second kinds.
Consequently, the obtained analytical results allow us to investigate the asymmetrical fermion system in the small and large gap parameter regimes, respectively.
In the limit of small pairing gap, we discuss and confirm on the issues of how the asymmetry in fermion species affects the phases of the ground state \cite{caldas}.
In the limit of large pairing gap, we find that there are two candidates competing for the ground state.
Depending on the degree of asymmetry, it can be either a pure Fermi fluid plus a mixed condensate or a pure Fermi fluid plus a pure boson condensate.
Moreover, it in principle enables us to interpolate the system in both regimes, too.

The plan of the paper is as follows.
In Section \ref{Model} we introduce the model Hamiltonian and derive a number of mean-field equations.
In Section \ref{ASI} and \ref{ASII} we solve these mean-field equations exactly at the zero temperature when two species of fermions have unequal number densities or chemical potentials.
In Section \ref{SIII} we analyze these equations with two species of fermions having the same number density.
We conclude the paper in Section \ref{CR}.

\section{\la{Model} The Model and Mean-Field Equations}

Let us consider a system of two species of fermions interacting through a short-range attractive potential, {\it i.e.}, a dilute system with asymmetrical fermion species.
The first species, denoted by $A$, has mass $M_A$ and total particle number $N_A$.
The other species, denoted by $a$, has mass $M_a$ and total number $N_a$.
Without loss of generality, we set \( M_A = m \, M_a \) with \( m \ge 1 \) and \( N_A = n \, N_a \) while keeping $n$ arbitrary ($0 < n < \infty$) in the discussions.
The nonrelativistic version of the system is well described by the grand canonical Hamiltonian
\be
{\cal H} = \sum_\p
\left( \,
\xi_p^A \, A^\dagger_\p \, A_\p +
\xi_p^a \, a^\dagger_{-\p} \, a_{-\p} \, \right) +
g \sum_{\p , \q} A^\dagger_{\p} \, a^\dagger_{-\p} \, a_{-\q} \, A_{\q} \, ,
\la{Hamiltonian1}
\ee
where $\p$ and $\q$ are momenta, $A_\p$, $a_\p$, $A^\dagger_\p$ and $a^\dagger_\p$ are the annihilation and creation operators, $g$ is the bare coupling constant and taken to be negative, in addition
$\xi_p^{A,a}$ = \( \varepsilon_{A,a} - \mu_{A,a} \) stand for the dispersion relations of the $A$-type and $a$-type fermions, respectively.
Here, \( \varepsilon_{A,a} = p^2/2M_{A,a} \) are the kinetic energies and $\mu_{A,a}$ are the chemical potentials.
The notation $p = | \, \p \, |$ is adopted.

Following the mean-field analysis of the BCS theory, we introduce in the Hamiltonian (\ref{Hamiltonian1}) a real order parameter, the pairing gap, as \( \Delta = - g \sum_\p \, \langle a_{-\p} \, A_\p \rangle \, \).
This allows us to simplify the Hamiltonian that can be further diagonalized via the Bogoliubov-Valatin transformation \cite{tinkham}.
It turns out that the diagonalized Hamiltonian is of the form
\be
{\cal H} = \sum_\p \left(
E_p^B \, B^\dagger_\p \, B_\p +
E_p^b \, b^\dagger_{-\p} \, b_{-\p} \right) +
\sum_\p \left( \xi_p^a - E_p^b \right) - \frac{\Delta^2}{g} \, .
\la{Hamiltonian2}
\ee
Here, $B_\p$, $b_\p$, $B^\dagger_\p$, and $b^\dagger_\p$ are annihilation and creation operators of the so-called $B$-type and $b$-type quasi-fermions, whose corresponding energies are denoted by $E_p^B$ and $E_p^b$, respectively.
The quasi-fermions $B$ and $b$ are mixture of the original fermions $A$ and $a$ and play the roles of the quasihole and quasiparticle excitations to the ground state of the mean-field Hamiltonian.
Explicitly, we have the operator transformation \( B^\dagger_\p = u_p \, A^\dagger_\p \, - \, v_p \, a_{-\p} \) and
\( b_{-\p} = u_p \, a_{-\p} \, + \, v_p \, A^\dagger_\p \), with the transformation functions $u_p$ and $v_p$ fulfill the relations:
\be
u_p^2 + v_p^2 = 1
\qquad {\rm and} \qquad
u_p^2 - v_p^2 = \frac{(\xi_p^A+\xi_p^a)}{\sqrt{(\xi_p^A+\xi_p^a)^2 + 4 \Delta^2}} \, .
\la{upvp}
\ee
In Eq. (\ref{Hamiltonian2}), the quasiparticle energies $E_p^B$ and $E_p^b$ are found to be
\ba
E_p^B &=& \frac{1}{2}
\left[ \, ( \xi_p^A - \xi_p^a ) + \sqrt{(\xi_p^A + \xi_p^a)^2 + 4 \Delta^2} \, \right]
\, ,
\la{EpB}
\\
E_p^b &=& \frac{1}{2}
\left[ \, - ( \xi_p^A - \xi_p^a ) + \sqrt{(\xi_p^A + \xi_p^a)^2 + 4 \Delta^2} \, \right]
\, .
\la{Epb}
\ea

If we designate the ground state of the system by $| \Psi \rangle$ \cite{wilczek.1,caldas}, at a finite temperature $T$,
the quasi-fermions will have occupation numbers described by
\( \langle \Psi | \, B^\dagger_p \, B_\p \, | \Psi \rangle = f_F (E_p^B) \)
and
\( \langle \Psi | \, b^\dagger_{-\p} \, b_{-\p} \, | \Psi \rangle = f_F (E_p^b) \), respectively.
Here, the Fermi-Dirac distribution function is \( f_F (E) = \frac{1}{e^{\beta E}+1} \), with \( \beta = \frac{1}{k_B T} \).
Therefore, it is straightforward to derive the thermodynamic potential $\Omega$ of the system at the ground state
$| \Psi \rangle$
\ba
\Omega (\Delta,\mu_A,\mu_a)
&=& \langle \Psi | \, {\cal H} \, | \Psi \rangle
\nonumber
\\
&=& \int \frac{d^3p}{(2\pi)^3}
\left[
f_F (E_p^B) \, E_p^B + f_F(E_p^b) \, E_p^b+ \xi_p^a - E_p^b
\right] -
\frac{\Delta^2}{g} \, .
\la{Omega}
\ea

The mean particle number density of each type of fermions $A$ and $a$ can be carried out by applying the thermodynamic property of the $\Omega (\Delta,\mu_A,\mu_a)$ as

\ba
N_A
&=& -\frac{\partial}{\partial \mu_A} \Omega (\Delta,\mu_A,\mu_a)
\nonumber
\\
&=&
\int \frac{d^3p}{(2\pi)^3} \left[ v_p^2 \, (1-f_F(E_p^b)) + u_p^2 \, f_F(E_p^B) \right] \, ,
\la{NA}
\\
N_a
&=& -\frac{\partial}{\partial \mu_a} \Omega (\Delta,\mu_A,\mu_a)
\nonumber
\\
&=&
\int \frac{d^3p}{(2\pi)^3} \left[ v_p^2 \, (1-f_F(E_p^B)) + u_p^2 \, f_F(E_p^b) \right] \, .
\la{Na}
\ea
Furthermore, a consistent equation for the gap parameter $\Delta$ is obtained by
\ba
\Delta
& \equiv & - g \sum_\p \langle \Psi | \, a_{-\p} \, A_\p \, | \Psi \rangle
\nonumber
\\
&=&
- g_R \int \frac{d^3p}{(2\pi)^3}
\left[
u_p \, v_p \left( 1 - f_F(E_p^b) - f_F(E_p^B) \right) -
\frac{\Delta}{\varepsilon_A + \varepsilon_a}
\right] \, .
\la{Delta}
\ea
Because of the choice of contact pairing interaction in (\ref{Hamiltonian1}), the gap equation diverges at the ultraviolet limits.
Thus, we are lead to introduce a temperature-independent-regulator to regularize the gap equation in the second line of Eq. (\ref{Delta}).
The quantity $g_R$ is the renormalized coupling constant that is related to the two-body (S-wave) scattering length in the low-density approximation by \( a_s = \frac{g_R M_A}{2(m+1)\pi} \).

Besides the thermodynamic potential (\ref{Omega}), we can as well perform the Legendre transformation to find the Helmholtz free energy ${\cal F}$ of the system
\be
{\cal F} (\Delta,N_A,N_a)
= \Omega (\Delta,\mu_A,\mu_a) + \mu_A N_A + \mu_a N_a \, ,
\la{Helmholtz}
\ee
where the chemical potentials $\mu_A$ and $\mu_a$ are presumably calculated in terms of the mean particle numbers $N_A$ and $N_a$ by inverting Eqs. (\ref{NA}) and (\ref{Na}).

At the zero temperature, the system is still properly described by the above equations (\ref{Omega}-\ref{Helmholtz}),
however, with the Fermi-Dirac function takes simple numbers.
Depending on the value of energy $E$, we have \( f_F(E) = 1 \) for $E < 0$ and \( f_F(E) = 0 \) for $E > 0$.
With the explicit expressions of (\ref{EpB}) and (\ref{Epb}), the zeros of quasiparticle energies $E_p^B$ and $E_p^b$ are located at the energy values
\be
\varepsilon_{\pm} = \frac{1}{2}
\left[ \,
( \mu_A + \bar\mu_a ) \pm
\sqrt{ ( \mu_A - \bar\mu_a )^2 - 4 \bar\Delta^2 } \,
\right] \, .
\la{epm}
\ee
For the purpose of simplicity, we have introduced in this equation the scaled chemical potential of $a$-type fermions as
\( \bar\mu_a \equiv \frac{\mu_a}{m} \) and the scaled gap parameter as \( \bar\Delta \equiv \frac{\Delta}{\sqrt{m}} \).

Both quasiparticle energies $E_p^B$ and $E_p^b$ are positive definite when $\varepsilon_{\pm}$ defined above (\ref{epm}) are found to be complex conjugate to each other.
However, when $\varepsilon_{+}$ and $\varepsilon_{-}$ are both real, then one of the quasiparticle energies, either $E_p^B$ or $E_p^b$, can take negative values in the energy interval
\( \varepsilon \in \left[ \varepsilon_{-}, \varepsilon_{+} \right] \).
Note that \( \varepsilon_{+} \ge \varepsilon_{-} \) and $\varepsilon_{-}$ may become negative.

We are now in a position to investigate the analytical results of the physical quantities, such as $\mu_A$, $\bar\mu_a$, $g_R$, $\Omega$, and ${\cal F}$ introduced above, when the system is at the ground state at the zero temperature.
We shall discuss three possibilities of interest below.
They are separately the asymmetrical superfluid with $N_A > N_a$, the asymmetrical superfluid with $N_A < N_a$, and
the symmetrical superfluid with $N_A = N_a$.

\section{\la{ASI} Asymmetrical Superfluid with $N_A > N_a$}

The system is in one branch of the so-called Sarma phase (the BP phase), which can accommodate at the same time a Fermi liquid and a superfluid in the weak-coupling regime.
In this branch and thereinafter named the first branch, for a given value of gap parameter $\Delta$, the quasiparticle energy $E_p^B$ will become negative in the energy domain \( ( \varepsilon_{-} < \varepsilon < \varepsilon_{+} ) \), whereas $E_p^b$ is still kept positive.
From Eq. (\ref{epm}), it is shown that the allowed values of $\mu_A$ and $\bar\mu_a$ reside in the interval
\( \bar\mu_a - \mu_A \in ( - \infty, - 2 \bar\Delta ] \).
Note that \( \bar\mu_a = \frac{\mu_a}{m} \) and \( \bar\Delta = \frac{\Delta}{\sqrt{m}} \).
When \( \bar\mu_a - \mu_A = - 2 \bar\Delta \), the boundary points of the energy interval match, that is,
$\varepsilon_{-} = \varepsilon_{+} \).
Moreover, the lower bound of the interval $\varepsilon_{-} > 0$ for
\( \bar\mu_a \in ( - \frac{\bar\Delta^2}{\mu_A} , \mu_A - 2 \bar\Delta ) \)
and $\varepsilon_{-} < 0$ for \( \bar\mu_a \in ( - \infty, - \frac{\bar\Delta^2}{\mu_A} ) \).
Based on the analysis, the thermodynamic potential (\ref{Omega}) can be reformulated into
\ba
&& \Omega^{(1)} (\Delta_m,\mu_A,\mu_a)
\nonumber
\\
&&
\quad = \frac{(2 M_A)^{3/2}}{4 \pi^2}
\Bigg\{
\frac{m+1}{2}
\left( \int_0^{\varepsilon_{-}} + \int_{\varepsilon_{+}}^\infty \right)
d \varepsilon \sqrt{\varepsilon}
\left[
\left( \varepsilon - \mu_T \right) -
\frac{(\varepsilon - \mu_T)^2 + 2 \Delta_m^2}{\sqrt{(\varepsilon - \mu_T)^2 +
(2 \Delta_m)^2}}
\right]
\nonumber
\\
&&
\quad + \int_{\varepsilon_{-}}^{\varepsilon_{+}} d \varepsilon
\sqrt{\varepsilon} \left( \varepsilon - \mu_A \right)
\Bigg\}
\, ,
\la{OmegaI}
\ea
where the kinetic energy \( \varepsilon \equiv \varepsilon_A = \frac{p^2}{2M_A}\) of the $A$-type fermions is used as the integration variable.
The new notation for the gap parameter is defined by
\( \Delta_m \equiv \frac{\Delta}{m+1} = \frac{\sqrt{m}}{m+1} \bar\Delta \)
and the total chemical potential is \( \mu_T \equiv \frac{\mu_A+\mu_a}{m+1} = \frac{\mu_A+m {\bar\mu}_a}{m+1} \).
Here, the superscript $^{(1)}$ of the $\Omega$ implies that we are in the first branch of Sarma phase, such that
\( N_A > N_a \) (see Eq. (\ref{NANaI}) and the paragraph below) or, equivalently, \( \mu_A - \bar\mu_a \ge 2 \bar\Delta \).

In deriving Eq. (\ref{OmegaI}), we have combined the gap parameter equation (\ref{Delta}) with the thermodynamic potential Eq. (\ref{Omega}) for the elimination of the bare coupling $g$.
Choosing the same integration variable $\varepsilon$, the gap equation (\ref{Delta}) becomes the integral equation for the renormalized coupling constant $g_R$, which takes the following form
\be
1 = - \frac{g_R}{(m+1)} \frac{M_A^{3/2}}{\sqrt{2} \pi^2}
\left\{
\left( \int_0^{\varepsilon_{-}} + \int_{\varepsilon_{+}}^\infty \right)
d \varepsilon \sqrt{\varepsilon}
\left[ \frac{1}{\sqrt{(\varepsilon - \mu_T)^2 + (2 \Delta_m)^2}} -
\frac{1}{\varepsilon} \right] -
\int_{\varepsilon_{-}}^{\varepsilon_{+}}
\frac{d \varepsilon }{\sqrt{\varepsilon}}
\right\}
\, .
\la{DeltaI}
\ee

In the same manner, the particle number density equations, Eqs. (\ref{NA}) and (\ref{Na}), are reshuffled into
\ba
&& N_A - N_a = \frac{M_A^{3/2}}{2\sqrt{2} \pi^2} \frac{4}{3}
\left[ (\varepsilon_{+})^{3/2} - (\varepsilon_{-})^{3/2} \right] \, ,
\la{NANaI}
\\
&& N_a = \frac{M_A^{3/2}}{2\sqrt{2} \pi^2}
\left( \int_0^{\varepsilon_{-}} + \int_{\varepsilon_{+}}^\infty \right)
d \varepsilon \sqrt{\varepsilon}
\left[ 1-\frac{\varepsilon - \mu_T}{\sqrt{(\varepsilon - \mu_T)^2 +
(2 \Delta_m)^2}} \right] \, .
\la{NaI}
\ea
Here, we obtain an algebraic equation plus an integral equation in Eqs. (\ref{NANaI}) and (\ref{NaI}), instead of the original two integral equations given in Eqs. (\ref{NA}) and (\ref{Na}).
Since \( \varepsilon_{+} > \varepsilon_{-} \), in Eq. (\ref{NANaI}) the number of $A$-type particles has to exceed that of $a$-type particles.
This validates our earlier claim.
That is to say, some of $A$-type particles will remain in the normal state even after the pairing mechanism occurs.
As a result, we expect that the chemical potential $\mu_A$ of $A$-type fermions will stay in the positive region and never penetrate the zero value for any finite gap parameter $\Delta_m$.

At zero pairing gap \( \Delta_m = 0 \), the chemical potentials for both species of fermions are easily calculated.
It is found that \( \mu_A (\Delta_m = 0) = \varepsilon^A_F \) and
\( \bar\mu_a (\Delta_m = 0) = (\frac{1}{n})^{2/3} \varepsilon^A_F \), where
\( \varepsilon^A_F = \frac{(6 \pi^2 N_A)^{2/3}}{2M_A} \) is the {\it free} Fermi energy of the $A$-type fermions, that is, the unpaired Fermi energy.
Here, \( n = \frac{N_A}{N_a} > 1 \) is the particle number ratio.
It is pointing out that when the combined factor \( m (\frac{1}{n})^{2/3} > 1 \) with the mass ratio
\( m = \frac{M_A}{M_a} \), then \( \mu_a = m \bar\mu_a > \mu_A \) at $\Delta_m = 0$.
As the gap parameter increases under the circumstances of fixing $n$ a constant, the value of $\mu_a$ will decrease in response and eventually go to negative region, signifying the formation of molecular bound state.
However, the value of $\mu_A$ should remain in the positive region.
We thus expect a crossing between $\mu_A$ and $\mu_a$ curves that will happen at some finite value of $\Delta_m$.
The particular solutions of chemical potentials so that \( \mu_A (\Delta_m) = \mu_a (\Delta_m) \) can be found by solving the number equation (\ref{NANaI}).
Generically, it is an algebraic equation of degree six and will simplify to an equation of degree four for the choice of mass ratio \( m=1 \).
On the contrary, when the combined factor \( m (\frac{1}{n})^{2/3} < 1 \), we should observe no crossing between the two chemical potential curves $\mu_A$ and $\mu_a$ to occur at any value of $\Delta_m$.

All the equations (\ref{OmegaI})-({\ref{NaI}) shown above are valid only for the case when $\varepsilon_{-} > 0$.
If we notice at some gap parameter $\Delta_m$ that $\varepsilon_{-} < 0$, then these equations need to be modified by simply letting $\varepsilon_{-} = 0$.
It is because that negative kinetic energy is not physically allowed.

To continue on the discussion of the analytical results, it is convenient to use dimensionless quantities.
For this purpose, we follow and adopt closely the conventions and notations of the reference \cite{marini}.
Let us introduce the dimensionless energy variables and dimensionless total chemical potential by
\be
\alpha_\pm = \frac{\varepsilon_\pm}{2 \Delta_m}
\qquad {\rm and} \qquad
x_0 = \frac{\mu_T}{2 \Delta_m}
\, ,
\la{dimensionless}
\ee
and then express Eqs. (\ref{OmegaI})-(\ref{NaI}) using them.

In the sequel, the analytical expressions can be obtained and will be presented in following paragraphs.
It is found that, in terms of the dimensionless quantities $\alpha_{\pm}$ and $x_0$, the thermodynamic potential
(\ref{OmegaI}) becomes
\ba
&& \Omega^{(1)} (\Delta_m,\mu_A,\mu_a)
\nonumber
\\
&& \quad =
\frac{m+1}{4 \pi^2}  \frac{(4 M_A \Delta_m)^{5/2}}{2 M_A}
\left[
\, \omega^{(1)} (x_0,0) + \omega^{(1)} (x_0,\alpha_{+}) - \omega^{(1)} (x_0,\alpha_{-}) \,
\right]
\, .
\la{OmegaI.1}
\ea
Similarly, the gap equation (\ref{DeltaI}) reduces to
\be
1 = - \frac{g_R M_A}{(m+1) \pi^2} (4 M_A \Delta_m)^{1/2}
\Big[ \, I_1 (x_0,0) + I_1 (x_0,\alpha_{+}) - I_1 (x_0,\alpha_{-} \, ) -
\left( \sqrt{\alpha_{+}} - \sqrt{\alpha_{-}} \right) \, \Big]
\, ,
\la{DeltaI.1}
\ee
and the number density equations (\ref{NANaI}) and (\ref{NaI}) are expressible as
\ba
&& N_A - N_a = \frac{1}{6 \pi^2} (4 M_A \Delta_m)^{3/2}
\left[ \, (\alpha_{+})^{3/2} - (\alpha_{-})^{3/2} \, \right]
\, ,
\la{NANaI.1}
\\
&& N_a = \frac{1}{4 \pi^2} (4 M_A \Delta_m)^{3/2}
\Big[ \, I_2 (x_0,0) + I_2(x_0,\alpha_{+}) - I_2(x_0,\alpha_{-}) \, \Big]
\, .
\la{NaI.1}
\ea
Moreover, the substitution of Eqs. (\ref{OmegaI.1}), (\ref{NANaI.1}), and (\ref{NaI.1}) into the Helmholtz free energy (\ref{Helmholtz}) renders
\ba
&& {\cal F}^{(1)} (\Delta_m, \mu_A, \mu_a)
\nonumber
\\
&&
\quad =
\frac{m+1}{4 \pi^2} \frac{(4 M_A \Delta_m)^{5/2}}{2 M_A}
\left[
\, f^{(1)} (x_0,0) + f^{(1)} (x_0,\alpha_{+}) - f^{(1)} (x_0,\alpha_{-}) \,
\right]
\, .
\la{HelmholtzI.1}
\ea
Because the Helmholtz free energy is a function of number density, we have in mind in this equation that
$\mu_A = \mu_A (N_A,N_a)$ and $\mu_a = \mu_a (N_A,N_a)$ and the relations are provided in Eqs. (\ref{NANaI.1}) and
(\ref{NaI.1}).

The functions $I_1 (x_0,\alpha)$ and $I_2 (x_0,\alpha)$ appearing in Eqs. (\ref{DeltaI.1}) and (\ref{NaI.1}) are integral functions and given by
\ba
I_1 (x_0, \alpha)
&=&
\int_{\sqrt{\alpha}}^\infty dx \, x^2
\left[ \frac{1}{\sqrt{(x^2 - x_0)^2 + 1}} - \frac{1}{x^2} \right]
\, ,
\la{I1}
\\
I_2 (x_0, \alpha)
&=&
\int_{\sqrt{\alpha}}^\infty d x \, x^2
\left[ 1 - \frac{x^2 - x_0}{\sqrt{(x^2 - x_0)^2 + 1}} \right]
\, ,
\la{I2}
\ea
where \( x^2 = \frac{\varepsilon}{2 \Delta_m} \) is dimensionless.
Also, the functions $\omega^{(1)} (x_0,\alpha)$ and $f^{(1)} (x_0,\alpha)$ in Eqs. (\ref{OmegaI.1}) and
(\ref{HelmholtzI.1}) are short-hand notation of the combinations
\ba
\omega^{(1)} (x_0, \alpha)
&=&
I_7 (x_0, \alpha) - \frac{1}{2} I_1 (x_0, \alpha) - x_0 I_2 (x_0, \alpha) +
\frac{2}{m+1}
\left[
\frac{\alpha^{5/2}}{5} - \frac{\alpha^{3/2}}{3} \frac{\mu_A}{2 \Delta_m}
\right] ,
\la{omegaI}
\\
f^{(1)} (x_0, \alpha)
&=& I_7 (x_0, \alpha) - \frac{1}{2} I_1 (x_0, \alpha) +
\frac{2}{m+1} \frac{\alpha^{5/2}}{5}
\, ,
\la{fI}
\ea
where the function $I_7 (x_0, \alpha)$ above is defined as
\be
I_7 (x_0, \alpha)
= \int_{\sqrt{\alpha}}^\infty d x \, x^4
\left[
1 - \frac{x^2 - x_0}{\sqrt{(x^2 - x_0)^2 + 1}} - \frac{1}{2 x^4}
\right]
\, .
\la{I7}
\ee
It is stressed that this integral is well defined at the ultraviolet limits.

The detailed calculations of the integral functions $I_1 (x_0,\alpha)$, $I_2 (x_0,\alpha)$, and $I_7 (x_0,\alpha)$ are straightforward but somewhat tedious.
The derivation is similar to that reported in the reference \cite{marini}.
Therefore, we only present the results and skip the cumbersome parts.
They can, respectively, be performed and expressed as linear combinations of the elliptic integrals of the first kind
$F(\phi,\kappa)$ and second kind $E(\phi,\kappa)$.
To be precise, we obtain the following exact results
\be
I_1 (x_0, \alpha) =
\sqrt{\alpha}
\left( 1 - \frac{\sqrt{(\alpha-x_0)^2 + 1}}{\alpha + \sqrt{1 + x_0^2}} \right) +
(1+x_0^2)^{1/4}
\left( \frac{1}{2} F(\phi_\alpha, \kappa) - E(\phi_\alpha, \kappa) \right)
\, ,
\la{I1.1}
\ee
\ba
I_2 (x_0, \alpha)
&=&
- \frac{\sqrt{\alpha}}{3}
\left[
\alpha - \sqrt{(\alpha-x_0)^2+1}
\left( 1 + \frac{x_0}{\alpha+\sqrt{1+x_0^2}} \right)
\right]
\nonumber
\\
&+&
\frac{1}{6} (1+x_0^2)^{1/4}
\left[ \left( \sqrt{1+x_0^2} - x_0 \right) F(\phi_\alpha, \kappa) +
2 x_0 E(\phi_\alpha, \kappa) \right]
\, ,
\la{I2.1}
\ea
\ba
I_7 (x_0, \alpha)
&=&
\frac{\sqrt{\alpha}}{10}
\left[
5 - 2 \alpha^2 + 2 \sqrt{(\alpha-x_0)^2+1}
\left( \alpha + x_0 + \frac{x_0^2-3}{\alpha+\sqrt{1+x_0^2}} \right)
\right]
\nonumber
\\
&+&
\frac{1}{10} (1+x_0^2)^{1/4}
\left[ \left( 3 - x_0^2 + x_0 \sqrt{1+x_0^2} \right) F(\phi_\alpha, \kappa) +
2 (x_0^2-3) E(\phi_\alpha, \kappa) \right]
\, .
\la{I7.1}
\ea
Here, the angle $\phi_\alpha$ and the modulus $\kappa$ appearing in the arguments of the elliptic integrals
$F(\phi,\kappa)$ and $E(\phi,\kappa)$ are defined by
\be
\phi_\alpha = \cos^{-1}
\left( \, \frac{\alpha - \sqrt{1+x_0^2}}{\alpha + \sqrt{1+x_0^2}} \, \right)
\, ,
\la{phi}
\ee
and
\be
\kappa^2 = \frac{1}{2}
\left( 1 + \frac{x_0}{\sqrt{1+x_0^2}} \right)
\, .
\la{kappa}
\ee
Note that $\kappa^2 < 1$.
Now we can compare the formulas of $I_1 (x_0, \alpha)$ and $I_2 (x_0, \alpha)$ to those calculated in the paper
\cite{marini}.
When letting $\alpha = 0$ so that $\phi_\alpha = \pi$, it is found that $I_1 (x_0, 0)$ and $I_2 (x_0, 0)$ given in Eqs. (\ref{I1}) and (\ref{I2}) reproduce exactly the desired answers.

Besides the analytical results, it is straightforward to check the approximate expressions of our results in the limits, when $x_0 \to \pm \infty$.
This can be achieved by simply retaining a few significant terms in the corresponding expansions.

We will analyze and present these approximate results in the case that the particle numbers $N_A$ and $N_a$ are both fixed quantities.
Hence, the particle number ratio $n$ is also fixed.
The discussion here is relevant to the system of trapped cold fermionic atoms, where the atoms are confined in a real space.
A remark regarding to the case when the total particle number is a fixed quantity is also illustrated.

The first limit of interest is when $x_0 \gg 1$ so that
\( \kappa^2 \simeq 1 - \frac{1}{4 x_0^2} \) = \( 1 - \left( \frac{\Delta_m}{\mu_T} \right) ^2 \).
Thus, it indicates the small gap approximation \( \Delta_m \to 0 \), or simply it means \( \Delta_m \ll \varepsilon_F^A \).
We define three constants by $A_0 = 1$, $a_0 = \left( \frac{1}{n} \right)^{2/3}$ and \( T_0 = \frac{A_0 + m a_0}{m+1}  \).
Under such definition, the constants satisfy the inequality $a_0 < T_0 < A_0$.
Then, via the Eqs. (\ref{NANaI.1}) and (\ref{NaI.1}), the chemical potentials $\mu_A^{(1)}$ and $\bar\mu_a^{(1)}$ are found to have the following serial expansions
\be
\mu_A^{(1)} (\Delta_m) =
\varepsilon_F^A
\left[ A_0 + \frac{t}{\sqrt{A_0}}
\left(
\frac{m+1}{\sqrt{A_0} - \sqrt{a_0}} - \frac{{\cal G} (A_0,a_0)}{2 \sqrt{T_0}}
\right) + {\cal O} (t^2)
\right]
\, ,
\la{muAI}
\ee
\be
\bar\mu_a^{(1)} (\Delta_m) =
\varepsilon_F^A
\left[ a_0 - \frac{t}{\sqrt{a_0}}
\left(
\frac{m+1}{m ( \sqrt{A_0} - \sqrt{a_0} \, )} + \frac{{\cal G} (A_0,a_0)}{2 \sqrt{T_0}}
\right) + {\cal O} (t^2)
\right]
\, ,
\la{muaI}
\ee
where \( t = \Big( \frac{\Delta_m}{\varepsilon_F^A} \Big)^2 \) is the expansion parameter and the ${\cal G}$-function is defined by
\be
{\cal G} (A_0,a_0) =
\ln
\left|
\frac{\sqrt{T_0} + \sqrt{A_0}}{\sqrt{T_0} - \sqrt{A_0}} \,
\frac{\sqrt{T_0}+\sqrt{a_0}}{\sqrt{T_0}-\sqrt{a_0}}
\right|
\, .
\la{Gfunction}
\ee
In Eqs. (\ref{muAI}) and (\ref{muaI}), we observe in the limit $\Delta_m \ll \varepsilon_F^A$ that
\( \mu^{(1)}_A (\Delta_m) - \mu^{(1)}_A (0) > 0 \) and \( \bar\mu^{(1)}_a (\Delta_m) - \bar\mu^{(1)}_a (0) < 0 \).
Both the chemical potentials are adjusted accordingly to accommodate the change of the band structure due to the nonzero pairing gap.

Similarly, the gap equation (\ref{DeltaI.1}) turns out to be
\ba
\frac{1}{g_R^{(1)} (\Delta_m)}
&=&
\frac{M_A k_F^A}{(m+1) \pi^2}
\Bigg\{
\sqrt{A_0} + \sqrt{a_0} - \frac{\sqrt{T_0}}{2} \, {\cal G} (A_0,a_0)
\nonumber
\\
&+&
\frac{t}{8 T_0 \sqrt{a_0 A_0}}
\Bigg[
\frac{2 (1+m)}{m}
\frac{(\sqrt{A_0}+m\sqrt{a_0}\,)(2T_0-\sqrt{a_0 A_0}\,)}{(\sqrt{A_0}-\sqrt{a_0}\,)^2}
\nonumber
\\
&-& \frac{\sqrt{a_0 A_0} - 4T_0}{\sqrt{T_0}} \,
{\cal G} (A_0,a_0) +
\frac{\sqrt{a_0} + m\sqrt{A_0}}{m+1} \, {\cal G} (A_0,a_0)^2
\Bigg] +
{\cal O} (t^2)
\Bigg\} ,
\la{gRI}
\ea
where $k_F^A = \sqrt{2 M_A \varepsilon_F^A} = (6 \pi^2 N_A)^{1/3}$ is the {\it free} Fermi momentum of the $A$-type fermions at $\Delta_m =0$.
Recall that $\varepsilon_F^A = \frac{(k_F^A)^2}{2 M_A}$ is the associate {\it free} Fermi energy.

It is apparent that the solution of Eq. (\ref{gRI}) can exist only for \( \frac{1}{g_R} - \frac{1}{g_c} > 0 \), where we have suppressed the superscript $^{(1)}$ in the $g_R$.
The critical coupling constant $g_c$ is \cite{wilczek.1,caldas}
\be
\frac{1}{g_c} =
\frac{M_A k_F^A}{(m+1) \pi^2}
\left(
\sqrt{A_0} + \sqrt{a_0} - \frac{\sqrt{T_0}}{2} \, {\cal G} (A_0,a_0)
\right)
\, ,
\la{criticalgR}
\ee
and can take positive or negative values entirely determined by the sign of the object in the parenthesis.
In Essence, the sign of $g_c$ depends on the complicate combination of $n$ and $m$ values, and has
$\frac{1}{g_c} \to - \infty$ as the number ratio $n$ approaches unity.

Furthermore, the thermodynamic potential (\ref{OmegaI.1}) has the $t$-expanded series (in the limit $\Delta_m \ll \varepsilon_F^A$)
\ba
\Omega^{(1)} (\Delta_m)
&=&
- \frac{(k_F^A)^5}{4 \pi^2 M_A}
\Bigg[
\frac{2}{15} \left( A_0^{5/2} + m \, a_0^{5/2} \right)
\nonumber
\\
&+&
\frac{(m+1) \, t}{3}
\left(
\sqrt{A_0} + \sqrt{a_0} - \frac{\sqrt{T_0}}{2} \, {\cal G} (A_0,a_0)
\right) + {\cal O} (t^2)
\Bigg]
\, .
\la{T-potentialI}
\ea
From the expression of the $\Omega^{(1)} (\Delta_m)$ in Eq. (\ref{T-potentialI}), we arrive at an interesting conclusion concerning the status of the normal state in the grand canonical ensemble.
That is that the normal state is found stable, a global minimum, when the critical coupling  $g_c < 0$, whereas it becomes unstable, a local maximum, when $g_c > 0$.
Numerically, if we choose the mass ratio $m=2$, then the critical coupling $g_c$ reverses its sign at $n \approx 4.341$.
Moreover, the Helmholtz free energy (\ref{HelmholtzI.1}) is given by
\be
{\cal F}^{(1)} (\Delta_m) =
\frac{(k_F^A)^5}{4 \pi^2 M_A}
\left[
\frac{1}{5}
\left( A_0^{5/2} + m \, a_0^{5/2} \right) +
{\cal O} (t^2)
\right] .
\la{F-energyI}
\ee
When setting $t = 0$, Eq. (\ref{F-energyI}) is nothing but the Helmholtz free energy added separately from the unpaired $A$-type and unpaired $a$-type fermions.

In the same limit $\Delta_m \ll \varepsilon_F^A$, another important case to discuss is when the total particle number $(N_A + N_a)$ is kept fixed, but the conversion between $A$-type and $a$-type particle numbers is allowed.
This case is worth studying in the high energy physics.
The overall quark number is protected by the conservation of baryonic number, but quarks themselves can interchange their flavors due to the weak interaction.
In this case, we can easily extract the approximate results from Eq. (\ref{muAI}) to Eq. (\ref{F-energyI}) via the following simple replacements:
the Fermi momentum $k_F^A$ is replaced by $k_F^T = (6 \pi^2 (N_A+N_a))^{1/3}$,
the Fermi energy $\varepsilon_F^A$ is replaced by $\varepsilon_F^T$ = $\frac{(k_F^T)^2}{2 M_A}$,
the expansion parameter $t$ is replaced by $t'$ = $\Big( \frac{\Delta_m}{\varepsilon_F^T} \Big)^2$,
the constant $A_0$ is replaced by $A'_0 = \left( \frac{n}{n+1} \right)^{2/3}$,
the constant $a_0$ is replaced by $a'_0 = \left( \frac{1}{n+1} \right)^{2/3}$,
and lastly the constant $T_0$ is replaced by $T'_0 = \frac{A'_0 + m a'_0}{m+1}$.
For example, the chemical potential $\mu_A^{(1)} (\Delta_m)$ under which $(N_A+N_a)$ is a fixed constant has the $t'$-expansion
\be
\mu_A^{(1)} (\Delta_m) =
\varepsilon_F^T
\left[ A'_0 + \frac{t'}{\sqrt{A'_0}}
\left(
\frac{m+1}{\sqrt{A'_0} - \sqrt{a'_0}} - \frac{{\cal G} (A'_0,a'_0)}{2 \sqrt{T'_0}}
\right) + {\cal O} ({t'}^2)
\right]
\, .
\la{muAI-1}
\ee
All other quantities can be obtained by the analogous procedure.

Now supposed that we want to study how the status of the normal state evolves in the grand canonical ensemble, using the analytical result (\ref{OmegaI.1}), as we vary the particle number difference \( (N_A-N_a) \) = \( N_a (n-1) \) and keep the total particle number \( (N_A+N_a) \) = constant.
We see the following facts.
At the beginning, for a small value of $(n-1)$, the normal state is just a local minimum and there exists a global minimum at some value of $\Delta_m \ne 0$.
As the $(n-1)$ value increases, the depth of normal state grows deeper in the $( \Omega^{(1)} ~{\rm vs.}~ \Delta_m )$ plot.
Then, at some particular value of $(n-1)$, the system exhibits coexisting double minima, where both of them are stable.
Numerically, it happens roughly around $n \approx 1.23$ with $m=2$.
At this stage, the spatially homogeneous Sarma phase is unstable with respect to the decay into a spatially inhomogeneous mixed phase \cite{bedaque,caldas}.
If $(n-1)$ gets larger and larger, the normal state now replaces the former one and becomes the unique global minimum of the system.
Nevertheless, for even much larger values of $(n-1)$, the normal state will turn out to be unstable eventually.
It becomes a local maximum.
Using the $t'$-expanded version of the thermodynamic potential $\Omega^{(1)}$ (\ref{T-potentialI}), the unstable point is found numerically at $n \approx 4.431$ for $m=2$.

To this end, let us discuss the other extreme of interest that is defined by the limits $x_0 < 0$ and $|x_0| \gg 1$, so that \( \kappa^2 \simeq \frac{1}{4 x_0^2} \) = \( \left( \frac{\Delta_m}{\mu_T} \right)^2 \) is a very small quantity.
Equivalently, it stands for the situation that \( (\varepsilon_F^A)^2 \ll \Delta_m^2 \ll \mu_T^2 \).
In such limit, $\varepsilon_{-} < 0$ always holds, we therefore need to set $\varepsilon_{-} = 0$ in the derivation of the approximate results.
In the equations to be presented, we denote two constants by
${\tilde A}_0 = \left( 1 - \frac{1}{n} \right)^{2/3}$ and
${\tilde a}_0 = \left( \frac{3 n \pi}{4} \right)^2$,
then use
\( \frac{1}{t} = \Big( \frac{\varepsilon_F^A}{\Delta_m} \Big)^2 \) as the expansion parameter.

It is found that the chemical potentials in the limit \( x_0 \to -\infty \) are
\be
\mu_A^{(1)} (\Delta_m) =
\varepsilon_F^A
\left[ {\tilde A}_0 + \frac{1+m}{{\tilde a}_0} \frac{1}{t} + {\cal O} ( t^{-3} )
\right]
\, ,
\la{hmuAI}
\ee
\be
\bar\mu_a^{(1)} (\Delta_m) = -
\varepsilon_F^A \, t^2
\left[
\frac{m+1}{m} {\tilde a}_0 + \frac{{\tilde A}_0}{m} \frac{1}{t^2} + {\cal O} ( t^{-3} )
\right]
\, .
\la{hmuaI}
\ee
Also, the gap equation can be written by
\be
\frac{1}{g_R^{(1)} (\Delta_m)} =
\frac{M_A k_F^A}{(m+1) \pi^2}
\left[ \frac{\pi}{2} \sqrt{{\tilde a}_0} \, t +
\left(
\frac{7 {\tilde A}_0^{3/2}}{3 \, {\tilde a}_0} -
\frac{13 \pi}{16 \, {\tilde a}_0^{3/2}}
\right) \frac{1}{t^2} +
{\cal O} ( t^{-3} )
\right]
\, .
\la{hgRI}
\ee
Hence, the system in the weak positive coupling regime $g_R \to 0^+$ is characterized by the limit of large pairing gap
$\Delta_m \gg \varepsilon_F^A$.
It is interesting to note the implication of the above equations.
Since the renormalized coupling constant $g_R$ is expressible in terms of the two-body scattering length as
\( g_R = \frac{2 (m+1) \pi}{M_A} \, a_s \), we are able to verify using Eqs. (\ref{hmuaI}) and (\ref{hgRI}) that
\( \mu_a^{(1)} = m \bar\mu_a^{(1)} \simeq - \frac{m+1}{2 M_A a_s^2} \) is just the binding energy of the two-particle, $A$ and $a$, bound state.
In addition, from Eq. (\ref{hmuAI}), the $\mu_A^{(1)} \simeq {\tilde A}_0 \, \varepsilon_F^A$ is the chemical potential of leftover $A$-type fermions that in fact become noninteracting after the formation of the tightly bound diatomic molecules.
Note that the number of the leftover $A$-type fermions is
\( (N_A - N_a) \) = \( N_A \Big( 1- \frac{1}{n} \Big) \) = \( N_A {\tilde A}_0^{3/2} \).

The thermodynamic potential admits the approximate form
\be
\Omega^{(1)} (\Delta_m) = -
\frac{(m+1) (k_F^A)^5}{8 \pi^2 M_A}
\Bigg[
\frac{4 {\tilde A}_0^{5/2}}{15(m+1)} -
\left(
\frac{2 {\tilde A}_0^{3/2}}{5 \, {\tilde a}_0} -
\frac{17 \pi}{40 \, {\tilde a}_0^{3/2}}
\right) \frac{1}{t} +
{\cal O} (t^{-3})
\Bigg]
\, .
\la{hT-potentialI}
\ee
Since the thermodynamic potential is related to the negative of the pressure of the system, Eq. (\ref{hT-potentialI}) shows that the excess $A$-type fermions contribute mainly to the system's pressure.
This makes sense because all the bound-state molecules are condensate in the ground state, thus they barely affect the pressure of the system.

Finally, the Helmholtz free energy is
\be
{\cal F}^{(1)} (\Delta_m) =
- \frac{(m+1) (k_F^A)^5}{8 \pi^2 M_A}
\left[
\frac{\pi}{2} \sqrt{{\tilde a}_0} \, t^2 -
\frac{2 {\tilde A}_0^{5/2}}{5(m+1)} +
\left(
\frac{44 {\tilde A}_0^{3/2}}{15 \, {\tilde a}_0} -
\frac{29 \pi}{20 \, {\tilde a}_0^{3/2}}
\right) \frac{1}{t} +
{\cal O} (t^{-3})
\right]
\, .
\la{hF-energyI}
\ee
In the limit $\Delta_m \gg \varepsilon_F^A$, it receives contributions primarily from the tightly bound $A$-$a$ molecular condensate and secondarily from the leftover $A$-type fermions.

Once more, if we want to express the equations (\ref{hmuAI}) to (\ref{hF-energyI}) under the situation of fixed overall particle number, \( (N_A+N_a) \) = constant, then we only need in every equation to replace
$k_F^A$ by $k_F^T = (6 \pi^2 (N_A+N_a))^{1/3}$,
$\varepsilon_F^A$ by $\varepsilon_F^T$,
$t$ by $t'$,
${\tilde A}_0$ by ${\tilde A'}_0 = \left( \frac{n-1}{n+1} \right)^{2/3}$, and finally
${\tilde a}_0$ by ${\tilde a'}_0 = \Big( \frac{3(n+1)\pi}{4} \Big)^2$.
Let us remind that \( n = \frac{N_A}{N_a} > 1 \) in the first branch of the Sarma phase.

\section{\la{ASII} Asymmetrical Superfluid with $N_A < N_a$}

The system is in the other branch of the Sarma phase, thus will be called the second branch of the Sarma phase.
For a nonzero value of gap parameter $\Delta$, the quasiparticle energy $E_p^b$ may be negative in the energy range
\( \varepsilon_{-} \le \varepsilon \le \varepsilon_{+} \), while $E_p^B$ is still positive definite.
Using Eq. (\ref{epm}), one can show that the chemical potential difference is given by
\( \bar\mu_a - \mu_A \in [2 \bar\Delta, \infty ) \).
The equality $\varepsilon_{-} = \varepsilon_{+}$ is fulfilled when \( \bar\mu_a - \mu_A = 2 \bar\Delta \).
Furthermore, we have $\varepsilon_{-} > 0$
for \( \mu_A \in ( - \frac{\bar\Delta^2}{\bar\mu_a} , \bar\mu_a - 2 \bar\Delta ) \) and
$\varepsilon_{-} < 0$ for \( \mu_A \in ( - \infty, - \frac{\bar\Delta^2}{\bar\mu_a} ) \).

In the second branch, the analytical results can be carried out in a straightforward manner that is similar to what we have presented in the case of the first branch, that is, the situation of $N_A > N_a$.
We find that the analytical results of the number density for $A$-type and $a$-type fermions, Eqs. (\ref{NA}) and
(\ref{Na}), are arranged into the forms
\ba
&& N_a - N_A = \frac{1}{6 \pi^2} (4 M_A \Delta_m)^{3/2}
\left[ \, (\alpha_{+})^{3/2} - (\alpha_{-})^{3/2} \, \right]
\, ,
\la{NANaII}
\\
&& N_A = \frac{1}{4 \pi^2} (4 M_A \Delta_m)^{3/2}
\Big[ \, I_2 (x_0,0) + I_2(x_0,\alpha_{+}) - I_2(x_0,\alpha_{-}) \, \Big]
\, ,
\la{NAII}
\ea
where the dimensionless quantities $x_0$ and $\alpha_\pm$ are defined in Eq. (\ref{dimensionless}) and
\( \Delta_m = \frac{\Delta}{m+1} \).

Since \( \alpha_{+} > \alpha_{-} \), we conclude that \( N_a > N_A \) in the second branch.
That is, some of the $a$-type fermions will stay in the normal state after pairing occurs.
The both chemical potentials at zero gap parameter are easily derived as
\( \mu_A (\Delta_m = 0) = \varepsilon^A_F \) and \( \bar\mu_a (\Delta_m = 0) = (\frac{1}{n})^{2/3} \varepsilon^A_F \).
Here, \( n = \frac{N_A}{N_a} < 1 \).
Since the combination \( m (\frac{1}{n})^{2/3} > 1 \) is always true in the second branch, we see that \( \mu_a > \mu_A \) at $\Delta_m = 0$.
As the gap parameter increases under the constraint that $n$ is kept fixed, the value of $\mu_A$ will decrease and eventually go to negative region, while $\mu_a$ (or $\bar\mu_a$) continues to stay in positive region and never across the zero value.
Thus, we do not expect any crossing between the chemical potential curves $\mu_A$ and $\mu_a$ to happen at any value of $\Delta_m$.

The analytical result for the pairing gap equation in the second branch is shown to take the identical expression as that in the first branch.
That is, the gap equation in this branch is nothing but Eq. (\ref{DeltaI.1}).
As for the thermodynamic potential (\ref{Omega}), it is of this form
\ba
&& \Omega^{(2)} (\Delta_m,\mu_A,\mu_a)
\nonumber
\\
&& \quad =
\frac{m+1}{4 \pi^2} \frac{(4 M_A \Delta_m)^{5/2}}{2 M_A}
\left[
\, \omega^{(2)} (x_0,0) + \omega^{(2)} (x_0,\alpha_{+}) - \omega^{(2)} (x_0,\alpha_{-}) \,
\right]
\, .
\la{OmegaII}
\ea
Here, the superscript $^{(2)}$ of $\Omega$ means that we are dealing with what is defined in the second branch of the Sarma phase, which implies \( N_A < N_a \) or, equivalently, \( \mu_A - \bar\mu_a \le - 2 \bar\Delta \).
The Helmholtz free energy (\ref{Helmholtz}) is simplified to
\ba
&& {\cal F}^{(2)} (\Delta_m, \mu_A, \mu_a)
\nonumber
\\
&&
\quad =
\frac{m+1}{4 \pi^2} \frac{(4 M_A \Delta_m)^{5/2}}{2 M_A}
\left[
\, f^{(2)} (x_0,0) + f^{(2)} (x_0,\alpha_{+}) - f^{(2)} (x_0,\alpha_{-}) \,
\right]
\, .
\la{HelmholtzII}
\ea
Because the Helmholtz free energy is a function of number density, we have to solve for $\mu_A = \mu_A (N_A,N_a)$ and $\mu_a = \mu_a (N_A,N_a)$ using in Eqs. (\ref{NANaII}) and (\ref{NAII}).

Both functions $\omega^{(2)} (x_0, \alpha)$ and $f^{(2)} (x_0, \alpha)$ shown in (\ref{OmegaII}) and (\ref{HelmholtzII}) are of the combinations
\ba
\omega^{(2)} (x_0, \alpha)
&=&
I_7 (x_0, \alpha) - \frac{1}{2} I_1 (x_0, \alpha) - x_0 I_2 (x_0, \alpha) +
\frac{2m}{m+1}
\left[
\frac{\alpha^{5/2}}{5} - \frac{\alpha^{3/2}}{3} \frac{\bar\mu_a}{2 \Delta_m}
\right] ,
\la{omegaII}
\\
f^{(2)} (x_0, \alpha)
&=& I_7 (x_0, \alpha) - \frac{1}{2} I_1 (x_0, \alpha) +
\frac{2m}{m+1} \frac{\alpha^{5/2}}{5}
\, .
\la{fII}
\ea
Recall that the exact results of $I$-functions, $I_1 (x_0, \alpha)$, $I_2 (x_0, \alpha)$, and $I_7 (x_0, \alpha)$, are given in Eqs. (\ref{I1.1}), (\ref{I2.1}), and (\ref{I7.1}), respectively

All the equations presented above are valid only when $\varepsilon_{-} > 0$.
For the case when $\varepsilon_{-} < 0$, the correct equations are obtained by setting $\varepsilon_{-} = 0$ in the corresponding ones.

Now we want to investigate the approximate results in the second branch, in the limits when $x_0 \to \pm \infty$.
Again, we keep the number ratio $n$ a constant in the discussion, that is, both particle numbers $N_A$ and $N_a$ are separately fixed.
Then, we will give a brief remark regarding to the case of the overall number $(N_A+N_a)$ fixed.

The first interesting limit is when $x_0 \gg 1$, which signifies the small gap approximation \( \Delta_m \to 0 \) or $\Delta_m \ll \varepsilon_F^A$, equivalently.
With the help of the number equations (\ref{NANaII}) and (\ref{NAII}), we find that the serial expansions of the chemical potentials in the second branch turn out to have exactly the same forms as those found in the first branch.
Explicitly, we have $\mu_A^{(2)} = \mu_A^{(1)}$ and $\bar\mu_a^{(2)} = \bar\mu_a^{(1)}$.
See Eqs. (\ref{muAI}) and (\ref{muaI}) for the details.
As far as the other quantities are concerned, they all take the same approximate expressions in both branches of Sarma phase.
That is, the approximate coupling constant $g_R^{(2)}$ equals $g_R^{(1)}$ in Eq. (\ref{gRI}), the approximate thermodynamic potential $\Omega^{(2)}$ (\ref{OmegaII}) equals $\Omega^{(1)}$ in Eq. (\ref{T-potentialI}), and the approximate Helmholtz free energy ${\cal F}^{(2)}$ (\ref{HelmholtzII}) equals ${\cal F}^{(1)}$ in Eq. (\ref{F-energyI}).
In fact, in the limit $\Delta_m \ll \varepsilon_F^A$ the expansions of all the quantities in both branches match.
However, a minor difference to be mentioned is that in the second branch the constants $A_0$, $a_0$ and $T_0$ satisfy the inequality $a_0 > T_0 > A_0$, instead of that $a_0 < T_0 < A_0$ in the first branch.
Numerically, we find  in the second branch that the critical coupling $g_c$ reverses the sign at $n \approx 0.019$ for the choice of the mass ratio $m=2$.

With regard to the case of fixing total particle number $(N_A+N_a)$, all the relevant equations can be deduced by the simple replacements that have been addressed in the context of the paragraph of the equation (\ref{muAI-1}).

That the same question arises in the second branch, similar to that in the first branch, is how the normal state changes in the grand canonical ensemble, when the number difference $(N_a-N_A)$ varies and the overall number is kept fixed.
Choosing $m=2$ in the analytical expression of the thermodynamic potential (\ref{OmegaI.1}), we find that the normal state will remain as a local minimum and never become a global minimum for the value of number ratio ranged from $n=1$ to as small as $n=0.1$.
This implies that in the second branch the system can only have a chance to exhibit coexisting double minima at very large asymmetry in fermion species.
The specific value of $n$ showing this property is roughly around 0.08.
Again, if the value of $n$ gets even small and is smaller than 0.019, {\it i.e.}, the critical value that $g_c$ in the $t'$-expanded version reverses the sign, then the normal state will turn out forever to be a local maximum.

The other interesting limit is given by $x_0 < 0$ and $|x_0| \gg 1$, which implies that \( (\varepsilon_F^A)^2 \ll \Delta_m^2 \ll \mu_T^2 \).
Again, $\varepsilon_{-} < 0$ in this limit, so that we have to set $\varepsilon_{-} = 0$ in the derivation.
By denoting ${\hat A}_0 = \left( \frac{3 \pi}{4} \right)^2$ and ${\hat a}_0 = \left( \frac{1}{n} - 1 \right)^{2/3}$,
we present the approximate results as follows.

It is found that the chemical potentials in the limit \( x_0 \to -\infty \) are
\be
\mu_A^{(2)} (\Delta_m) = -
\varepsilon_F^A \, t^2
\left[ (1+m) {\hat A}_0 + m \, {\hat a}_0 \frac{1}{t^2} + {\cal O} ( t^{-3} )
\right]
\, ,
\la{hmuAII}
\ee
\be
\bar\mu_a^{(2)} (\Delta_m) =
\varepsilon_F^A
\left[
{\hat a}_0 + \frac{(1+m)}{m {\hat A}_0} \frac{1}{t} + {\cal O} ( t^{-3} )
\right]
\, .
\la{hmuaII}
\ee
where \( \frac{1}{t} = \Big( \frac{\varepsilon_F^A}{\Delta_m} \Big)^2 \) is the expansion parameter.
Also, the gap equation, which signifies the system in the weak positive coupling regime $g_R \to 0^+$, can be written by
\be
\frac{1}{g_R^{(2)} (\Delta_m)} =
\frac{M_A k_F^A}{(m+1) \pi^2}
\left[ \frac{\pi}{2} \sqrt{{\hat A}_0} \, t +
\left(
\frac{7 \, {\hat a}_0^{3/2}}{3 {\hat A}_0} -
\frac{13 \pi}{16 {\hat A}_0^{3/2}}
\right) \frac{1}{t^2} +
{\cal O} ( t^{-3} )
\right]
\, .
\la{hgRII}
\ee
In terms of the two-body scattering length $a_s$, we find that \( \mu_A^{(2)} \simeq - \frac{m+1}{2 M_A a_s^2} \) is the binding energy of molecular bound state, and that
$\mu_a^{(2)} = m \bar\mu_a^{(2)} \simeq m \, {\hat a}_0 \, \varepsilon_F^A$ is the chemical potential of excess $a$-type fermions after the formation of the tightly bound molecules.
The excess number of $a$-type fermions is
\( N_a - N_A \) = \( N_A \Big( \frac{1}{n} -1  \Big) \) = \( N_A {\hat a}_0^{3/2} \).

In addition, the thermodynamic potential admits the $\frac{1}{t}$-expansion
\be
\Omega^{(2)} (\Delta_m) = -
\frac{(m+1) (k_F^A)^5}{8 \pi^2 M_A}
\Bigg[
\frac{4 m \, {\hat a}_0^{5/2}}{15(m+1)} -
\left(
\frac{2 \, {\hat a}_0^{3/2}}{5 {\hat A}_0} -
\frac{17 \pi}{40 {\hat A}_0^{3/2}}
\right) \frac{1}{t} +
{\cal O} (t^{-3})
\Bigg]
\, .
\la{hT-potentialII}
\ee
Eq. (\ref{hT-potentialII}) means that only the excess $a$-type fermions contribute to the system's pressure in the limit $\frac{1}{t} \to 0$, since $\Omega = - P$, the negative of the pressure.

At last, the Helmholtz free energy in the limit $\Delta_m \gg \varepsilon_F^A$ becomes
\be
{\cal F}^{(2)} (\Delta_m) =
- \frac{(m+1) (k_F^A)^5}{8 \pi^2 M_A}
\left[
\frac{\pi}{2} \sqrt{{\hat A}_0} \, t^2 -
\frac{2 m \, {\hat a}_0^{5/2}}{5(m+1)} +
\left(
\frac{44 \, {\hat a}_0^{3/2}}{15 {\hat A}_0} -
\frac{29 \pi}{20 {\hat A}_0^{3/2}}
\right) \frac{1}{t} +
{\cal O} (t^{-3})
\right] .
\la{hF-energyII}
\ee
Both the bound $A$-$a$ molecular condensate and the leftover $a$-type fermions contribution to the Helmholtz free energy.

As regards the resembling equations when the total particle number $(N_A+N_a)$ is fixed, the replacements below are found useful.
That is, $k_F^A$ by $k_F^T$, $\varepsilon_F^A$ by $\varepsilon_F^T$, $t$ by $t'$,
${\hat a}_0$ by ${\hat a'}_0 = \left( \frac{1-n}{1+n} \right)^{2/3}$, and at last
${\hat A}_0$ by ${\hat A'}_0 = \Big( \frac{3(n+1)\pi}{4 n} \Big)^2$.
A reminder is that $n = \frac{N_A}{N_a} <1$ in the second branch of the Sarma phase.

\section{\la{SIII} Symmetrical Superfluid with $N_A = N_a$}

The system is in the so-called BCS phase, in which the two species of fermions have the same particle number.
In this phase, both quasiparticle energies $E_p^B$ and $E_p^b$ are always positive for any given gap parameter $\Delta$.
That is, $E_p^B > 0$ and $E_p^b > 0$.
The allowed chemical potentials satisfy the relation \( \bar\mu_a - \mu_A \in [ -2 \bar\Delta, 2 \bar\Delta ] \).
The analytical results for the number density of $A$ and $a$ fermions (\ref{NA}) and (\ref{Na}) take the simple forms
\be
N_A = N_a = \frac{1}{4 \pi^2} (4 M_A \Delta_m)^{3/2} I_2 (x_0,0)
\, ,
\la{NANaIII}
\ee
where, as indicated, both $A$-type and $a$-type particles have the same number.
Here, the number ratio \( n = \frac{N_A}{N_a} = 1 \).
When the gap parameter vanishes, the integration of (\ref{NANaIII}) is easily performed.
It is found that \( \mu_A = \bar\mu_a = \varepsilon^A_F \) for \( \Delta_m = 0 \).
Recall that \( \Delta_m = \frac{\Delta}{m+1} \).

The analytical result for the pairing gap equation (\ref{Delta}) is written by
\be
1 = - \frac{g_R M_A}{(m+1) \pi^2} (4 M_A \Delta_m)^{1/2} I_1 (x_0,0)
\, .
\la{DeltaIII}
\ee
Furthermore, the thermodynamic potential (\ref{Omega}) and the Helmholtz free energy (\ref{Helmholtz}) are, respectively
\ba
\Omega^{(0)} (\Delta_m,\mu_T)
&=&
\frac{m+1}{4 \pi^2} \frac{(4 M_A \Delta_m)^{5/2}}{2 M_A}
\left[ I_7 (x_0, 0) - \frac{1}{2} I_1 (x_0, 0) - x_0 I_2 (x_0, 0) \right]
\, ,
\la{OmegaIII}
\\
{\cal F}^{(0)} (\Delta_m, \mu_T)
&=&
\frac{m+1}{4 \pi^2} \frac{(4 M_A \Delta_m)^{5/2}}{2 M_A}
\left[ I_7 (x_0, 0) - \frac{1}{2} I_1 (x_0, 0) \right]
\, ,
\la{HelmholtzIII}
\ea
where \( \mu_T = \frac{\mu_A + \mu_a}{m+1} \) is the system's chemical potential in the BCS phase.
The superscript $^{(0)}$ means that we are investigating the quantities of the system in the BCS phase, so that
\( N_A = N_a \) or, equivalently, \( \mu_A - \bar\mu_a \in [ \, -2 \bar\Delta, 2 \bar\Delta \, ] \).
The exact results of the $I_1 (x_0, 0)$, $I_2 (x_0, 0)$, and $I_7 (x_0, 0)$ functions are given in
Eqs. (\ref{I1.1}), (\ref{I2.1}), and (\ref{I7.1}), respectively.

If we choose the mass ratio parameter \( m =1 \), it is apparent that Eq. (\ref{NANaIII}) to Eq. (\ref{HelmholtzIII}) reproduce the Leggett's equations for the BCS-BEC crossover with $\mu_T$ identified as the system's chemical potential and $2 \Delta_m = \Delta$ as the system's gap parameter \cite{eagles,leggett,engelbrecht}.
For the increase of gap parameter, we expect the value of $\mu_T$ to decrease accordingly.
Then, at some particular  value of $\Delta$, the chemical potential $\mu_T$ will go to negative region signifying the occurrence of the Bose-Einstein condensation.

It is easy to obtain the approximate forms in the limits when $x_0 \to \pm \infty$.
The first limit to study is for the case $x_0 \gg 1$, so that $\Delta_m \to 0$ or $\Delta_m \ll \varepsilon_F^A$.
By solving Eq. (\ref{NANaIII}), the system's chemical potential $\mu_T^{(0)}$ has the serial expansion
\be
\mu_T^{(0)} (\Delta_m) =
\varepsilon_F^A
\left[ 1 - t
\left(
\ln \frac{4}{\sqrt{t}} + \frac{1}{2}
\right) + {\cal O} (t^2)
\right]
\, ,
\la{muTIII}
\ee
where \( t = \Big( \frac{\Delta_m}{\varepsilon_F^A} \Big)^2 \) is the expansion parameter.
Similarly, the gap equation Eq. (\ref{DeltaIII}) is
\be
\frac{1}{g_R^{(0)} (\Delta_m)} =
- \frac{M_A k_F^A}{(m+1) \pi^2}
\left[
\ln \frac{4}{\sqrt{t}} - 2 +
t
\left(
\ln \frac{4}{\sqrt{t}} - \frac{5}{4} -
\frac{1}{2} \Bigg( \ln \frac{4}{\sqrt{t}} \Bigg)^2
\right) +
{\cal O} (t^2)
\right] .
\la{gRIII}
\ee
This gives us the famous BCS value of the gap parameter at zero temperature,
\( \Delta_m \simeq 4 e^{-2} \, e^{- \pi / 2 k_F^A |a_s|} \, \varepsilon_F^A \).
It is clear that in the BCS phase there is always a solution to the gap equation, no matter how small the coupling constant $g_R$ is.
This is in contrast to the cases discussed in the Sarma phase.

Furthermore, in this limit $\Delta_m \ll \varepsilon_F^A$ the approximate expressions for the thermodynamic potential (\ref{OmegaIII}) and the Helmholtz free energy (\ref{HelmholtzIII}) reduce to
\be
\Omega^{(0)} (\Delta_m) =
- \frac{(m+1) (k_F^A)^5}{8 \pi^2 M_A}
\Bigg[
\frac{4}{15} -
\frac{2 t}{3}
\left(
\ln \frac{4}{\sqrt{t}} - 1
\right) + {\cal O} (t^2)
\Bigg]
\, ,
\la{T-potentialIII}
\ee
and
\be
{\cal F}^{(0)} (\Delta_m) =
\frac{(m+1) (k_F^A)^5}{8 \pi^2 M_A}
\left[
\frac{2}{5} - t +
{\cal O} (t^2)
\right]
\, .
\la{F-energyIII}
\ee
When setting $t = 0$, Eq. (\ref{F-energyIII}) is nothing but the Helmholtz free energy of the unpaired $A$-type and unpaired $a$-type fermions.

Using the approximate equations (\ref{F-energyI}) and (\ref{F-energyIII}), it is easy to check in the context of canonical ensemble that the inhomogeneous mixed phase is indeed the favored ground state over the homogeneous Sarma phase in the limit of small pairing gap.
For this purpose, let us consider a trapped system of two species of fermions $A$ and $a$ with separately fixed particle numbers $N_A$ and $N_a$.
As regards the Helmholtz free energies in the two branches of the Saram phase, they both take the same form as shown in Eq. (\ref{F-energyI}), that is, ${\cal F}^{(1)} (\Delta_m)$ = ${\cal F}^{(2)} (\Delta_m)$.
With respect to that in the inhomogeneous mixed phase, it contains a Fermi liquid state of unpaired fermions plus a fully gapped BCS state of paired fermions.
Explicitly, for the case of $n=\frac{N_A}{N_a} > 1$, the Helmholtz free energy in the inhomogeneous mixed phase is given by
\be
{\cal F}^{(1)}_{{\rm mix}} (\Delta_m) =
\frac{ \big( k_F^A {\tilde A}_0^{1/2} \big)^5}{20 \pi^2 M_A} +
\frac{(m+1) (k_F^A)^5}{8 \pi^2 M_A}
\Bigg[
\frac{2}{5} \, a_0^{5/2} - a_0^{1/2} \, t +
{\cal O} (t^{2})
\Bigg]
\, ,
\la{mix1}
\ee
where the first term on the left-handed side is from Fermi liquid and the second term is from BCS phase.
The surface tension effect of the interface is not taken into account in the analysis.
Note that ${\tilde A}_0 = \left( 1 - \frac{1}{n} \right)^{2/3}$ and $a_0 = \left( \frac{1}{n} \right)^{2/3}$.
In addition, for the case of $n = \frac{N_A}{N_a} < 1$, the Helmholtz free energy in the mixed phase is then
\be
{\cal F}^{(2)}_{{\rm mix}} (\Delta_m) =
\frac{ \big( k_F^A {\hat a}_0^{1/2} \big)^5}{20 \pi^2 M_A} +
\frac{(m+1) (k_F^A)^5}{8 \pi^2 M_A}
\Bigg[
\frac{2}{5} \, A_0^{5/2} - A_0^{1/2} \, t +
{\cal O} (t^{2})
\Bigg]
\, ,
\la{mix2}
\ee
where $A_0 =1$ and ${\hat a}_0 = \left( \frac{1}{n}-1 \right)^{2/3}$.
The direct comparsion of the Helmholtz free energies between the Sarma and mixed phases results in
\( ( {\cal F}^{(1)}-{\cal F}^{(1)}_{{\rm mix}} ) \sim ( A_0^{5/2}-a_0^{5/2}-{\tilde A}_0^{5/2} ) > 0 \) for the number ratio $n>1$ and
\( ( {\cal F}^{(2)}-{\cal F}^{(2)}_{{\rm mix}} ) \sim m ( a_0^{5/2}-A_0^{5/2}-{\hat a}_0^{5/2} ) > 0 \) for the number ratio $0<n<1$.
In both branches, the Sarma phase always has the higher Helmholtz free energy than the mixed phase does.
Therefore, in the limit \( \Delta_m \ll \varepsilon_F^A \) the spatially inhomogeneous mixed phase is found to be the energetically preferable ground state for the asymmetrical fermion system \cite{caldas}.

The other interesting limit is when $x_0 < 0$ and $|x_0| \gg 1$, that is, \( (\varepsilon_F^A)^2 \ll \Delta_m^2 \ll \mu_T^2 \).
By denoting $\tau_0 = \left( \frac{3 \pi}{4} \right)^2$, it is found that the chemical potential is
\be
\mu_T^{(0)} (\Delta_m) = -
\varepsilon_F^A \, t^2
\left[ \tau_0 - \frac{3}{\tau_0} \frac{1}{t^3} + {\cal O} ( t^{-4} )
\right]
\, ,
\la{hmuTIII}
\ee
where \( \frac{1}{t} = \Big( \frac{\varepsilon_F^A}{\Delta_m} \Big)^2 \) is the expansion parameter.
Also, the gap equation can be written by
\be
\frac{1}{g_R^{(0)} (\Delta_m)} =
\frac{M_A k_F^A}{(m+1) \pi^2}
\left[ \frac{\pi}{2} \sqrt{\tau_0} \, t -
\frac{5 \pi}{8 \, \tau_0^{3/2}}
\frac{1}{t^2} +
{\cal O} ( t^{-4} )
\right]
\, .
\la{hgRIII}
\ee
In terms of the two-body scattering length $a_s$, we obtain that \( \mu_T^{(0)} \simeq - \frac{m+1}{2 M_A a_s^2} \) is the binding energy of molecular bound state in the weak positive coupling regime.

Furthermore, the thermodynamic potential admits the $\frac{1}{t}$-expansion
\be
\Omega^{(0)} (\Delta_m) = -
\frac{(m+1) (k_F^A)^5}{8 \pi^2 M_A}
\left[
\frac{7 \pi}{20 \, \tau_0^{3/2}}
\frac{1}{t} +
{\cal O} (t^{-4})
\right]
\, .
\la{hT-potentialIII}
\ee
Eq. (\ref{hT-potentialIII}) indicates that the system's pressure is extremely small in the limit $\frac{1}{t} \to 0$.

At last, the Helmholtz free energy is
\be
{\cal F}^{(0)} (\Delta_m) =
- \frac{(m+1) (k_F^A)^5}{8 \pi^2 M_A}
\left[
\frac{\pi}{2} \sqrt{\tau_0} \, t^2 -
\frac{23 \pi}{20 \, \tau_0^{3/2}} \frac{1}{t} +
{\cal O} (t^{-4})
\right]
\, .
\la{hF-energyIII}
\ee
Here, it results solely from the tightly bound $A$-$a$ molecules, the standard Bose-Einstein (BE) condensate.

We end this section by posting a frequently asked question.
In the large pairing gap limit \( \Delta_m \gg \varepsilon_F^A \), what is the preferable ground state in the context of canonical ensemble for the trapped cold asymmetrical fermion system?
From the phase structures, we know that there are two competing candidates.
The first one is the Sarma phase consisting of a pure Fermi fluid and a mixed condensate, which is characterized by the homogeneous mixture of fermions in a boson condensate.
The other one is the separated phase composed of a pure Fermi fluid plus a standard BE condensate.
We resolve this problem by computing and comparing the Helmholtz free energies of the system at the two phases, based on the analytical results presented in this and previous two sections.

First, we consider a trapped system of $A$-type and $a$-type fermions with particle numbers \( N_A > N_a \).
The numbers $N_A$ and $N_a$ are both fixed quantities.
The Helmholtz free energy ${\cal F}^{(1)} (\Delta_m)$ in the Sarma phase is given by
Eq. (\ref{hF-energyI}).
In the separated phase, the system is made of a standard BE condensate of $N_a$ tightly bound $A$-$a$ molecules and a Fermi fluid of $(N_A-N_a)$ excess $A$-type fermions.
With the help of Eq. (\ref{hF-energyIII}), the Helmholtz free energy ${\cal F}^{(1)}_{{\rm sep}}$ in the separated phase is therefore
\be
{\cal F}^{(1)}_{{\rm sep}} (\Delta_m) =
\frac{ \big( k_F^A {\tilde A}_0^{1/2} \big)^5}{20 \pi^2 M_A} -
\frac{(m+1) (k_F^A)^5}{8 \pi^2 M_A}
\Bigg[
\frac{\pi}{2} \sqrt{{\tilde a}_0} \, t^2 -
\frac{23 \pi}{20 \, {\tilde a}_0^{3/2}}
\frac{1}{t} +
{\cal O} (t^{-4})
\Bigg]
\, ,
\la{sep1}
\ee
where the first term on the left-handed side is from Fermi liquid and the second term is from BE condensate.
To construct (\ref{sep1}), we neglect the surface tension effect of the interface between the two components.
Then, the subtraction of the Helmholtz free energies renders
\be
{\cal F}^{(1)} (\Delta_m) - {\cal F}^{(1)}_{{\rm sep}} (\Delta_m) \simeq -
\frac{(m+1) (k_F^A)^5}{8 \pi^2 M_A}
\Bigg[
\frac{44 {\tilde A}_0^{3/2}}{15 \, {\tilde a}_0} -
\frac{3 \pi}{10 \, {\tilde a}_0^{3/2}}
\Bigg]
\frac{1}{t}
\, .
\la{Dsep.1}
\ee
Hence, we have
\( \big( {\cal F}^{(1)} - {\cal F}^{(1)}_{{\rm sep}} \big) < 0 \) for the number ratio \( n > \frac{25}{22} \), and
\( \big( {\cal F}^{(1)} - {\cal F}^{(1)}_{{\rm sep}} \big) > 0 \) for \( 1 < n < \frac{25}{22} \).
Recall that ${\tilde a}_0 = \left( \frac{3 n \pi}{4} \right)^2$.
This implies that at large degree of asymmetry the system favors the Sarma phase as the ground state, while at small degree of asymmetry it prefers in the separated phase, the state containing a pure fermion region and a pure boson region.
This makes sense since the more amounts the excess fermions are in the trap, the better chance they have to be in the boson condensate due to the degeneracy pressure among them.
In addition, the Sarma phase and the separated phase are in fact degenerate in the limit $t \to \infty$.

Next, we discuss another trapped system of $A$-type and $a$-type fermions with particle numbers \( N_a > N_A \).
In the Sarma phase, the Helmholtz free energy ${\cal F}^{(2)} (\Delta_m)$ can be found in Eq. (\ref{hF-energyII}).
While, in the separated phase, the system is made of a BE condensate of $N_A$ tightly bound $A$-$a$ molecules and a Fermi fluid of $(N_a-N_A)$ excess $a$-type fermions.
Based on the Eq. (\ref{hF-energyIII}) in the BCS phase, the Helmholtz free energy ${\cal F}^{(2)}_{{\rm sep}}$ in the separated phase is now written by
\be
{\cal F}^{(2)}_{{\rm sep}} (\Delta_m) =
\frac{ m \, \big( k_F^A {\hat a}_0^{1/2} \big)^5}{20 \pi^2 M_A} -
\frac{(m+1) (k_F^A)^5}{8 \pi^2 M_A}
\Bigg[
\frac{\pi}{2} \sqrt{{\hat A}_0} \, t^2 -
\frac{23 \pi}{20 {\hat A}_0^{3/2}}
\frac{1}{t} +
{\cal O} (t^{-4})
\Bigg]
\, .
\la{sep2}
\ee
Similarly, we omit the surface tension effect on the interface.
Note that ${\hat A}_0 = \left( \frac{3 \pi}{4} \right)^2$.
Then, the subtraction of the ${\cal F}^{(2)}$ and ${\cal F}^{(2)}_{{\rm sep}}$ shows that
\be
{\cal F}^{(2)} (\Delta_m) - {\cal F}^{(2)}_{{\rm sep}} (\Delta_m) \simeq -
\frac{(m+1) (k_F^A)^5}{8 \pi^2 M_A}
\Bigg[
\frac{44 \, {\hat a}_0^{3/2}}{15 {\hat A}_0} -
\frac{3 \pi}{10 {\hat A}_0^{3/2}}
\Bigg]
\frac{1}{t}
\, .
\la{Dsep.2}
\ee
We have \( \big( {\cal F}^{(2)}-{\cal F}^{(2)}_{{\rm sep}} \big) < 0 \) for the number ratio \( 0 < n < \frac{22}{25} \), and \( \big( {\cal F}^{(2)} - {\cal F}^{(2)}_{{\rm sep}} \big) > 0 \) for \( \frac{22}{25} < n < 1 \).
The same previous finding applies to this second case.
In the large gap parameter limit, the Sarma phase supports the system at large degree of asymmetry, whereas the separated phase supports it at smaller asymmetry.
Both phases are degenerate at infinite pairing gap parameter.

\section{\la{CR} Concluding Remarks}

Based on the mean-field theory, we calculate and present the analytical results for the asymmetrical fermion system with attractive interaction at the zero temperature.
Generically, the results are found exact and expressible as linear combinations of the elliptic integrals of the first and second kinds.
We discuss in details the approximate expressions of the analytical results in two limits: one is at small gap parameter $\Delta_m \ll \varepsilon_F^A$ and the other is at large gap parameter $\Delta_m \gg \varepsilon_F^A$.
For the case of small pairing gap, we discuss how the normal state of free fermions evolves, depending on the asymmetry of fermion species.
Our results also indicate that the spatially homogeneous Sarma phase is indeed unstable with respect to the decay into the spatially inhomogeneous mixed phase.
In the limit of large pairing gap, we show that two candidate phases are competing for the ground state of the system.
The Sarma phase containing a pure Fermi fluid and a mixed condensate is favored at large degree of asymmetry.
The separated phase consisting of a pure Fermi fluid and a boson condensate supports the system at smaller degree of asymmetry.

In addition, from our calculation it is easy to demonstrate that the existence of the spatially inhomogeneous mixed phase is always possible.
Recall that the thermodynamic potential $\Omega$ is a function of the variables $\mu_A$, $\bar\mu_a$, and $\Delta_m$ and take different forms in different regions of chemical potentials.
In fact, the thermodynamic potential is of the combined expressions,
$\Omega$ = $\Omega^{(1)}$ (\ref{OmegaI.1}) in the region \( \mu_A - \bar\mu_a \ge 2 \bar\Delta \),
$\Omega$ = $\Omega^{(2)}$ (\ref{OmegaII}) in the region \( \mu_A - \bar\mu_a \le - 2 \bar\Delta \), and
$\Omega$ = $\Omega^{(0)}$ (\ref{OmegaIII}) in the region \( - 2 \bar\Delta \le \mu_A - \bar\mu_a \le 2 \bar\Delta \), respectively.
Therefore, if we plot the $( \Omega ~{\rm vs.}~ \bar\mu_a )$-diagram for various values of $\mu_A$ while keeping the gap parameter $\Delta_m$ fixed, we observe the followings. The profile of $\Omega$ is a continuous convcave-down function of the chemical potential $\bar\mu_a$, and is not differentiable precisely at the two points, \( \bar\mu_a = \mu_A - 2 \bar\Delta \) and \( \bar\mu_a = \mu_A + 2 \bar\Delta \).
As a result, the cone of tangents at each point describes various possible mixed phases that are composed of the inhomogeneous mixture of the normal Fermi liquid and the fully gapped BCS state \cite{wilczek.2}.

Despite we have provided in the paper the analytical results for the system of asymmetrical fermion species with attractive interaction, we intend not to address the issues relating to the system in the strong-coupling regime.
It is because that the Hamiltonian (\ref{Hamiltonian1}) requires to obtain the analytical results might be too simplified to describe the complicated physical systems.
To learn the system in the strong-coupling regime, one needs the more realistic Hamiltonian, such as at nonzero temperature, with finite-range interaction, and especially by the nonperturbative method, instead.

We conclude the paper by noting that we have studied the analytical results for a number of physical quantities defined at the mean-field level, but we have not yet evaluated other physical quantities that are defined beyond the mean-field theory.
For example, the introduction of Gaussian fluctuations on the mean-field background enables us to study the spatial fluctuations of the pairing gap parameter and the Goldstone mode of the broken symmetry \cite{marini,engelbrecht}.
We plan to investigate the analytical results for these physical quantities of the asymmetrical system in the future.

\vskip 1cm

The author is grateful to D. J. Han, C. R. Lee, F. K. Men, C. H. Pao, Y. C. Tsai, and S. T. Wu for useful and stimulating discussions.
This work was supported in part by Taiwan's National Science Council Grant No. 93-2112-M-194-010.

\end{document}